\documentclass[12pt]{article}

\usepackage{graphicx}
\usepackage{graphics}
\usepackage{epsfig}
\usepackage{amssymb}
\usepackage{amsmath}
\usepackage{color}
\usepackage{textcomp}
\usepackage{latexsym}
\usepackage{cite}

\usepackage[margin=2.5cm,footskip=0.6cm]{geometry}
\begin{document}

\begin{center}
{\Large{\bf Intruder Dynamics in a Frictional Granular Fluid: A Molecular Dynamics Study}} \\
\ \\
\ \\
by \\
Prasenjit Das$^{1,2}$, Sanjay Puri$^1$ and Moshe Schwartz$^{3}$ \\

$^1$School of Physical Sciences, Jawaharlal Nehru University, New Delhi 110067, India. \\
$^2$Department of Chemical and Biological Physics, Weizmann Institute of Science, Rehovot 76100, Israel. \\
$^3$Beverly and Raymond Sackler School of Physics and Astronomy, Tel Aviv University, Ramat Aviv 69934, Israel.
\end{center}

\begin{abstract}
We study the dynamics of an intruder moving through a fluidized granular medium in three dimensions ($d=3$). The intruder and grains have both translational and rotational degrees of freedom. The energy-dissipation mechanism is solid friction between all pairs of particles. We keep the granular system fluidized even at rather high densities by randomly perturbing the linear and angular velocities of the grains. We apply a constant external force of magnitude $F$ to the intruder, and obtain its steady state velocity $V_s$ in the center-of-mass frame of the grains. The $F$-$V_s$ relation is of great interest in the industrial processing of granular matter, and has been the subject of most experiments on this problem. We also obtain the mobility, which is proportional to the inverse viscosity, as a function of the volume fraction $\phi$. This is shown to diverge at the jamming volume fraction. For $\phi$ below the jamming fraction, we find that $V_s \sim F$ for small $F$ and $V_s \sim F^{1/2}$ for large $F$. The intruder shows diffusive motion in the plane perpendicular to the direction of the external force.
\end{abstract}

\newpage

\section{Introduction}
\label{sec1}

A granular material or powder consists of an assembly of solid particles or grains, which are polydispersed in size and shape and have many internal degrees of freedom~\cite{pg99,jn06,at06,jd94,bp04}. Granular materials show many unusual properties because of the dissipative interactions between the particles~\cite{gz93,gtz93,mus95,ums96,gr00,dp03,dap03,ap06,ap07,dps16,dps17}. In this context, the study of their flow properties is particularly important. Granular flow plays a significant role in many industrial applications such as transport of processed chemicals, pharmaceuticals, mineral ores, food stuff, powdered ceramics, and building materials. The flow properties of granular systems are quite different from those of ordinary liquids~\cite{pgp03,gm04,jfp06,fp08,gg14}. This is because kinetic energy in powders is dissipated via inelastic collisions or friction, and stored in intra-granular degrees of freedom~\cite{bp04,pb03}.

A significant problem in amorphous materials is the motion of an intruder through a system. A thorough study of the intruder motion helps us to understand the mechanical properties of various systems such as granular media~\cite{zsr92,cd09}, foams~\cite{deq05}, emulsions~\cite{hsl04}, suspensions or structural glasses~\cite{hor03}, etc. At high volume fractions, such materials show jamming and support a finite shear stress before yielding. Thus, a non-zero critical force is required to drive an intruder through such media.

The motion of an intruder in granular matter is a well-studied experimental problem. Unfortunately, the results from different experiments are not consistent. The experiment can be done in a {\it constant-force} ($F$) or a {\it constant-velocity} ($V$) configuration. In the first case, the intruder acquires a steady-state velocity $V_s$. In the second case, the intruder experiences an effective drag force $F_d$. Most studies focus on the drag of slow intruders in a dense granular medium~\cite{apb98,czw03,kcl04,gb04,gb05,RH2011,rfp11,gfp13,ht13}. (Of course, the volume fraction $\phi < \phi_J$, where $\phi_J$ is the jamming fraction. For a frictionless granular assembly, $\phi_J = 0.843$ in $d=2$, and $\phi_J = 0.639$ in $d=3$~\cite{ls10}.) In this limit, the intruder motion does not fluidize the granular material. In general, the observed results depend on the type of granular medium, density, intruder shape, boundary conditions, etc. Geng and Behringer~\cite{gb05} studied the drag force acting on an intruder in a $d=2$ granular material consisting of bi-disperse disks. They found that (a) $F_d$ shows a power-law dependence on the area fraction; and (b) $V \sim \exp(F_d)$. Further, the mobility of the intruder depends strongly on $V$~\cite{gb04}. Hilton and Tordesillas~\cite{ht13} showed that the drag force acting on a spherical intruder in a $d=3$ granular bed depends on the Froude number $Fr = 2V/\sqrt{gR}$, where $g$ is the gravitational acceleration, and $R$ is the radius of the intruder. For a frictional system with $Fr > 1$, $V \sim F_d$. For $Fr < 1$, they observe a deviation from the above linear behavior.

There exist very few studies of intruder dynamics in a dense granular medium in the high-velocity regime. The experimental study of Takehara et al.~\cite{tfo10} showed that $V \sim F_d^{1/2}$ for $\phi = 0.797$ in $d=2$. They also presented a scaling argument for this result. Later, Takehara and Okumura~\cite{to14} studied the drag force that acts on an intruder disk in $d=2$ with several values of $\phi \ge 0.76$. They confirmed the $V$-$F_d$ relation reported in Ref.~\cite{tfo10}. 

In constant-force experiments also, there is ambiguity about the precise $V_s$-$F$ relation in the low-velocity regime. Habdas et al.~\cite{hsl04} studied the motion of an intruder (magnetic bead) through a colloid in $d=3$ near the glass-transition volume fraction $\phi_g$. In their study, the $V_s$-$F$ relation becomes nonlinear as $\phi \rightarrow \phi_g$, viz., $V_s \sim F^3$. Hastings et al.~\cite{hor03} reported that $V_s \sim F^{3/2}$ for an intruder moving in a glassy background in $d=2$. The results in this regime probably depend on the nature of interaction forces among the particles and the dimensionality. Candelier and Dauchot~\cite{cd09} studied the creep motion of an intruder in a vibrated granular material close to jamming in $d=2$. They did not study the $V_s$-$F$ relationship.

There also exist a few simulation studies of $F_d$ acting on an intruder in a $d=2$ granular medium~\cite{wcz03,pug2012,bwz06,th16}. Bharadwaj et al.~\cite{bwz06} obtained $F_d$ for an immersed cylinder in a stream of solid particles. In their study, the drag force obeys the same relationship for both frictionless and frictional particles, i.e., $V \sim F_d^{1/2}$. More recently, Takada and Hayakawa~\cite{th16} studied $F_d$ acting on an intruder for both frictionless and frictional granular disks. They also considered cases with and without dry friction between the supporting base and the granular disks. In all cases, they reported that $V \sim F_d^{1/2}$.

To the best of our knowledge, there are no detailed simulations available for the intruder problem in $d=3$. However, most experiments and industrial applications are realized in $d=3$ geometries. In this paper, we address this gap in the literature, and numerically study the dynamics of an intruder in a fluidized granular medium for a wide range of volume fractions. Our primary goal in this paper is to obtain a systematic understanding of the $V$-$F$ relationship in both the low-velocity and high-velocity regimes.

We performed simulations in a constant-force configuration. In real experiments, the velocities of grains around the intruder are affected by the force gauge attached to it. In our simulations, no such complicating factors are present. The dissipative grain-grain and intruder-grain interactions are modeled by solid friction~\cite{dps16,bes10,sb11,dps17}. We focus on how the $V_s$-$F$ relation is affected by (a) the volume fraction of the grains, and (b) the relative diameter of the intruder particle vis-a-vis the grains. For low volume fractions and small $F$, we intuitively expect the relation between $V_s$ and $F$ to follow Stokes law: $V_s \sim F$. For higher volume fractions (but below $\phi_J$), experimental studies have been unclear about the precise form of the $V_s$-$F$ relationship. The main results of our numerical simulations and scaling arguments are as follows: \\
(a) At all values of $\phi$, $V_s \sim F^\beta$, with $\beta$ crossing over from $1$ to $1/2$ as $F$ increases. \\
(b) The inverse mobility of the intruder, which is proportional to the viscosity of the granular system, diverges as a power law of $(\phi_J - \phi)$ for $\phi \rightarrow \phi_J^{-}$. \\
(c) The intruder performs Brownian motion in the plane perpendicular to the direction of the external force.

This paper is organized as follows. In Sec.~\ref{sec2}, we present details of our modeling and simulations. We present comprehensive numerical results in Sec.~\ref{sec3}. Finally, we end this paper with a summary and discussion in Sec.~\ref{sec4}.

\section{Modeling and Numerical Details}
\label{sec2}

We use standard molecular dynamics (MD) techniques~\cite{at87,fs02,dr04} to simulate the motion of an intruder in a granular medium. The grains are identical: spherical in shape, and of equal mass $m$. Two particles with position vectors $\vec r_i$ and $\vec r_j$ interact via a two-body potential with a hard-core of diameter $R_1$, and a thin-shell repulsive potential of diameter $R_2$~\cite{dps16,dps17}. To be specific, we choose the interaction potential to be of the following form:
\begin{eqnarray}
\label{pot}
V(r) &=& \infty, \quad \quad r < R_1, \nonumber \\
&=& V_0 \left(\frac{R_2-r}{r-R_1}\right)^2, \quad \quad R_1 \le r < R_2, \nonumber \\
&=& 0, \quad \quad r \ge R_2 .
\end{eqnarray}
Here, $r=|\vec r_i - \vec r_j|$ is the separation between the two particles, $V_0$ is the amplitude of the potential, and $R_2 - R_1 < R_1$. Eq.~(\ref{pot}) models a repulsive potential which rises steeply from 0 at the outer boundary of the shell to infinity at the hard-core. The normal force acting on the $i^{\rm th}$ particle due to the  $j^{\rm th}$ particle is given by
\begin{eqnarray}
\label{fn}
\vec F_{ij}^n(r) = -\vec \nabla_i V(r),
\end{eqnarray}
where $\vec \nabla_i$ is the gradient with respect to $\vec r_i$. In earlier work, we have used this interaction potential to study freely evolving granular gases~\cite{dps16,dps17}, and heated granular systems in the low and high-density limits~\cite{dps18}. We confirmed numerically that the results were analogous to those for hard-sphere systems, Hertzian spheres and Hookeian spheres. Therefore, we believe that the usage of the interaction potential in Eq.~(\ref{pot}) does not introduce any artifacts in our simulation.

Let $(\vec v_i, \vec \omega_i)$ and $(\vec v_j, \vec \omega_j)$ denote the linear and angular velocities of the $i^{\rm th}$ and $j^{\rm th}$ particles, respectively. The velocity $\vec v_{ij}$ of the $i^{\rm th}$ particle relative to the  $j^{\rm th}$ particle at the effective touching point $(\vec r_i + \vec r_j)/2$ is given by
\begin{eqnarray}
\label{vrel}
\vec v_{ij} = \vec v_{i} - \vec v_{j} - \frac{1}{2}(\vec \omega_{i} + \vec \omega_{j})\times\vec r_{ij} ,
\end{eqnarray}
where $\vec r_{ij} = \vec r_i - \vec r_j$~\cite{daps17}. The corresponding solid friction force on the $i^{\rm th}$ particle is given by
\begin{eqnarray}
\label{ff}
\vec F_{ij}^f(r) = - \mu | \vec F_{ij}^n | \frac{\vec v_{ij}}{| \vec v_{ij}|},
\end{eqnarray}
where $\mu$ is the friction coefficient. In Eq.~(\ref{ff}), the frictional force $\vec F_{ij}^f$ has both tangential and normal components. (In principle, it is easy to remove the normal component but this reduces the computational efficiency. Further, the normal component does not play a significant role due to the stiffness of the radial potential. Eq.~(\ref{ff}) reduces to Coulombic friction when the thickness of the repulsive shell tends to zero. In that limit, our model reduces to a hard-sphere model where the relative velocity cannot have a normal component at the point of contact. Thus, $\vec F_{ij}^f$ becomes perpendicular to $\vec F_{ij}^n$~\cite{dps16,dps17}.) The torque on particle $i$ due to $j$ is given by
\begin{eqnarray}
\label{tork}
\vec \tau_{ij} = -\frac{1}{2}~\vec r_{ij} \times \vec F_{ij}^f.
\end{eqnarray}

We use the following units for various relevant quantities: lengths are expressed in units of $R_1$, energy in units of $u=V_0/10$, temperature in terms of $u/k_B$, and time as a multiple of $\sqrt{mR_1^2/V_0}$. For the sake of convenience and numerical stability, we set $R_1 = 1$, $R_2 = 1.1R_1$, $V_0 = 10$, $k_B = 1$, and $m = 1$. Therefore, the time unit is $t_0=1/\sqrt{10}$, and this allows us to use relatively large $\Delta t$ in our simulation. 

Next, we discuss the properties of the intruder, which is also spherical in shape. The diameter and mass of the intruder are, respectively, $R_{\rm int}=kR_1$ and $m_{\rm int}=k^3m_1$ ($k>1$), i.e., the mass densities of the grains and the intruder are the same. The diameter of the repulsive shell for the intruder is $k R_2$. The intruder-grain potential is the same as Eq.~(\ref{pot}) with $R_1 \rightarrow (1+k)R_1/2$ and $R_2 \rightarrow (1+k)R_2/2$. We apply a constant external force $F$ on the intruder along the $+x$-direction.

We perturb the linear and angular velocities of the grains at regular intervals, which keeps the granular medium in the fluidized phase~\cite{dps18,wm96,dw96}. Our modeling of the system is motivated by the experiment of Candelier and Dauchot~\cite{cd09}. The method of perturbing the grain velocities mimics the experiments of Ojha et al.~\cite{ojha04}, where translational and rotational velocities are perturbed in all directions. Then, the equations of motion for the $i^{\rm th}$ particle can be written as follows:
\begin{eqnarray}
\label{heatlin}
m\frac{d\vec v_i}{dt} & = & \vec F_i^{\,\rm tot} + \vec\eta_i,\\
\label{heatrot}
I\frac{d\vec \omega_i}{dt} & = & \vec \tau_i^{\,\rm tot} + \vec\sigma_i .
\end{eqnarray}
Here,
\begin{eqnarray}
\label{net_f}
\vec F_i^{\rm tot} & = & \sum_{j\ne i}\left(\vec F_{ij}^n + \vec F_{ij}^f\right) , \\
\label{net_t}
{\vec \tau_i}^{\rm tot} & = & \sum_{j\ne i} \vec{\tau}_{ij} ,
\end{eqnarray}
are the total force and torque acting on the $i^{\rm th}$ particle, respectively. We consider all the grains to be solid spheres, i.e., their moment of inertia is $I= mR_1^2/10$.

The noises $\vec\eta_i$ and $\vec\sigma_i$ in Eqs.~(\ref{heatlin})-(\ref{heatrot}) are uncorrelated, and satisfy the following relations:
\begin{eqnarray}
\label{noise1}
\langle \vec\eta_i\rangle &=& 0,\\
\label{noise2}
\langle \eta_{i,\alpha}(t) \eta_{j,\beta}(t')\rangle&=&m^2\xi^2\delta_{ij}\delta_{\alpha\beta}\delta(t-t'),\\
\label{noise3}
\langle \vec\sigma_i\rangle &=& 0,\\
\label{noise4}
\langle \sigma_{i,\alpha}(t) \sigma_{j,\beta}(t')\rangle&=&I^2\xi^2\delta_{ij}\delta_{\alpha\beta}\delta(t-t') .
\end{eqnarray}
Here, $\alpha, \beta = x, y, z$, and $\xi$ characterizes the strength of the stochastic force. The noises associated with the translational and rotational degrees of freedom stem from the same vibration of the system. The right-hand-side of Eqs.~(\ref{noise2}) and (\ref{noise4}) is chosen so that the kinetic temperatures associated with both noises are equal. During the simulations, we perturb the system after a time step $dt = m \Delta t$ ($m=200$), where $\Delta t$ is the integration time step. This is done by adding a random increment to the linear and angular velocities of each particle as
\begin{eqnarray}
\label{velupd}
v_{i,\alpha}(t+\Delta t) & = & v_{i,\alpha}(t) + \sqrt{\Gamma}\sqrt{dt}~\theta_{i,\alpha}, \\
\omega_{i,\alpha}(t+\Delta t) & = & \omega_{i,\alpha}(t) + \sqrt{\Gamma} \sqrt{dt}~\varphi_{i,\alpha} ,
\end{eqnarray}
where $\alpha = x, y, z$ and $\Gamma$ is the strength of the noise: $\Gamma = 12\xi^2$. The random numbers $\theta, \varphi$ are uniformly distributed in the interval $[-0.5, 0.5]$. We have confirmed through small-scale simulations that the results obtained for a uniform noise distribution are comparable to those for a Gaussian noise distribution.

The corresponding equations of motion for the intruder are analogous to Eqs.~(\ref{heatlin})-(\ref{heatrot}) but without the noise terms:
\begin{eqnarray}
\label{intlin}
m_{\rm int} \frac{d\vec v_{\rm int}}{dt} & = & \vec F_{\rm int}^{\,\rm tot}, \\
\label{introt}
I_{\rm int} \frac{d\vec \omega_{\rm int}}{dt} & = & \vec \tau_{\rm int}^{\,\rm tot} .
\end{eqnarray}

The details of our simulation are as follows. The velocity Verlet algorithm~\cite{at87,fs02,dr04} with the integration time step $\Delta t=0.0005$ is implemented to update the positions and velocities in the MD simulation. The granular system is confined to a $d=3$ box of size $70 \times 30 \times 30$. The longer side is in the $x$-direction, along which the force on the intruder acts. To obtain the desired volume fraction ($\phi$), we vary the number of grains ($N$) in the system as given in Table~\ref{table}. We apply periodic boundary conditions in all directions.
\begin{table}[h!]
	\begin{center}
		\caption{Number of grains ($N$)}
		\vskip 0.2cm
		\label{table}
		\begin{tabular}{|c|c|c|c|}
			\hline
			$R_{\rm int}$ & $\phi=0.10$ & $\phi=0.30$ & $\phi=0.45$\\
			\hline
			3$R_1$ & 12007 & 36070 & 55960\\
			\hline
			5$R_1$ & 11910 & 35972 & 55879\\
			\hline
		\end{tabular}
	\end{center}
\end{table}

The system is prepared at $t=-50$ by randomly placing the grains and the intruder in the simulation box, such that there is no overlap between the cores of any two particles. The grains are assigned the same speed, but the directions of velocity vectors are random so that $\sum_{i=1}^N \vec v_i = 0$. Clearly, this does not correspond to a Maxwell-Boltzmann (MB) distribution. The unperturbed ($\xi=0, F=0$) system is allowed to evolve without dissipation ($\mu = 0$) till $t = 0$. This elastic evolution relaxes the system to an MB velocity distribution and a uniform density field, as we have confirmed numerically.

At $t=0$, we also start with an MB distribution for angular velocities at the same ``temperature'' as the velocity distribution. This serves as the initial condition ($t=0$ state) for our simulation of inelastic spheres with $\mu \ne 0$, $\xi \ne 0$, and $F>0$ acting on the intruder along the $+x$-direction. We use $\mu=0.1$ and $\xi=1.0$. The results presented here are obtained as an average over 25 independent runs. In earlier work~\cite{dps18}, we have studied the temperature and the velocity distribution of the grains (without an intruder) with friction and a thermostat. At long times, the temperature settles to a near-constant value $T_s \sim \xi^{4/3}$, which can be understood by a simple scaling argument. Further, the steady-state velocity distribution is approximately Gaussian, with small departures which can be characterized by a Sonine polynomial expansion about the MB distribution.

\section{Detailed Numerical Results}
\label{sec3}
\begin{figure}
	\centering
	\includegraphics*[width=0.90\textwidth]{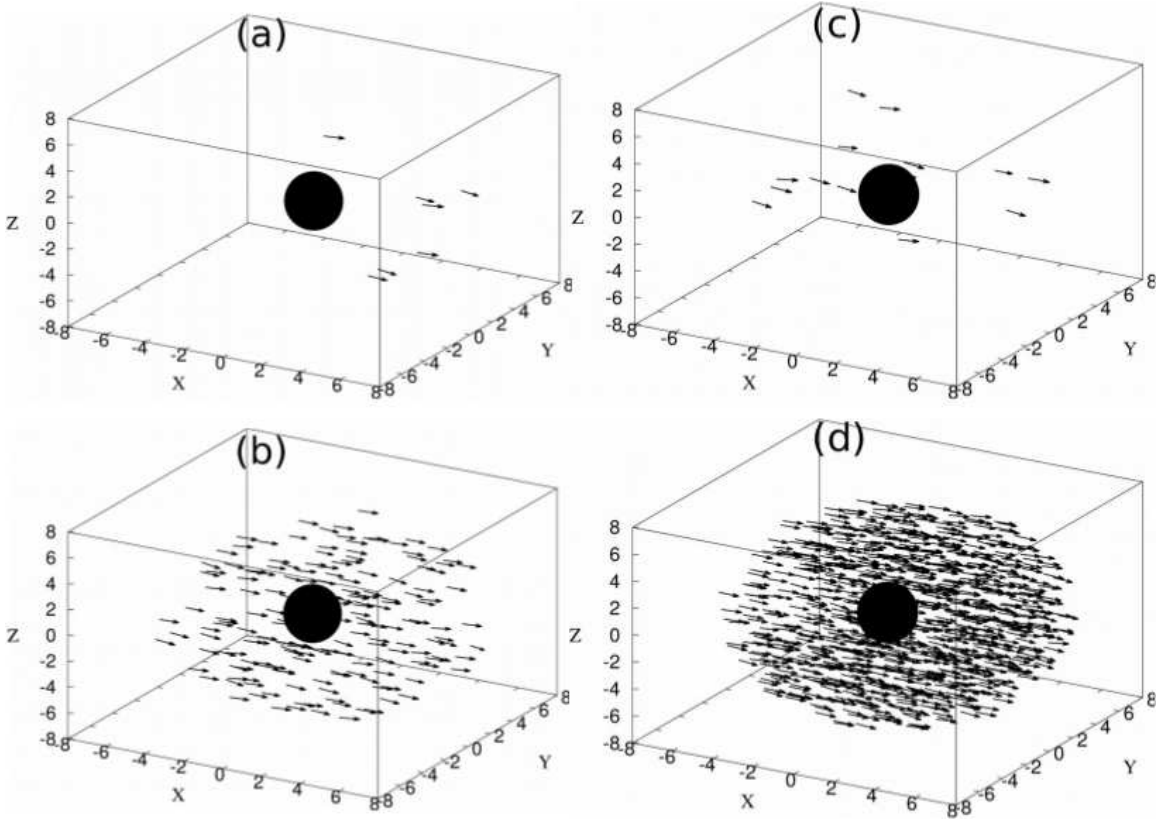}
	\caption{Linear velocities of the grains in the steady state, plotted in the lab reference frame. The external force on the intruder is $\vec F=F\hat{x}$. We show only aligned particles, whose velocities make an angle $< 0.2 \pi$ with the $+x$ direction, and which lie within a distance $7.5~R_1$ ($R_1=1$) from the center of the intruder. The center of the solid particle represents the position of the intruder, which is shifted to the origin. The starting point of a velocity vector represents the position of the center of a grain. The diameter of the intruder hard-core is $R_{\rm int}=3R_1$. The values of $F$ and $\phi$ are: (a) $F=10$, $\phi=0.10$, (b) $F=600$, $\phi=0.10$, (c) $F=10$, $\phi=0.30$, and (d) $F=600$, $\phi=0.30$.}
	\label{fig1}
\end{figure}

As discussed above, we start the simulation with a homogeneous density field. The linear and angular velocity fields are distributed via MB distributions with equal temperatures. At $t=0$, we apply a constant external force $F$ on the intruder along the $+x$ direction. A steady state is established by $t_s \simeq 100$ for all parameter values considered here. In Fig.~\ref{fig1}, we plot the linear velocities of the grains in the steady state. We show all grains whose centers lie within a distance $7.5R_1$ from the center of the intruder, and whose velocities make an angle $0 \rightarrow 0.2\pi$ with respect to the $+x$ direction. The lengths of the velocities are normalized to unity. For $F=10$ and $\phi=0.10$ [Fig.~\ref{fig1}(a)], most of the grains around the intruder have random velocities. We see that only a few grains have velocities aligned along the $+x$ direction. As we increase $F$, velocities of more grains become aligned along the direction of $F$, as shown in  Fig.~\ref{fig1}(b) for $F=600$ and $\phi=0.10$. For $\phi=0.30$, we also observe a similar behavior, as shown in Fig.~\ref{fig1}(c) and Fig.~\ref{fig1}(d) for $F=10$ and $F=600$, respectively. The alignment range of grain velocities vis-a-vis the intruder depends on both $F$ and $\phi$.

\begin{figure}
	\centering
	\includegraphics*[width=0.99\textwidth]{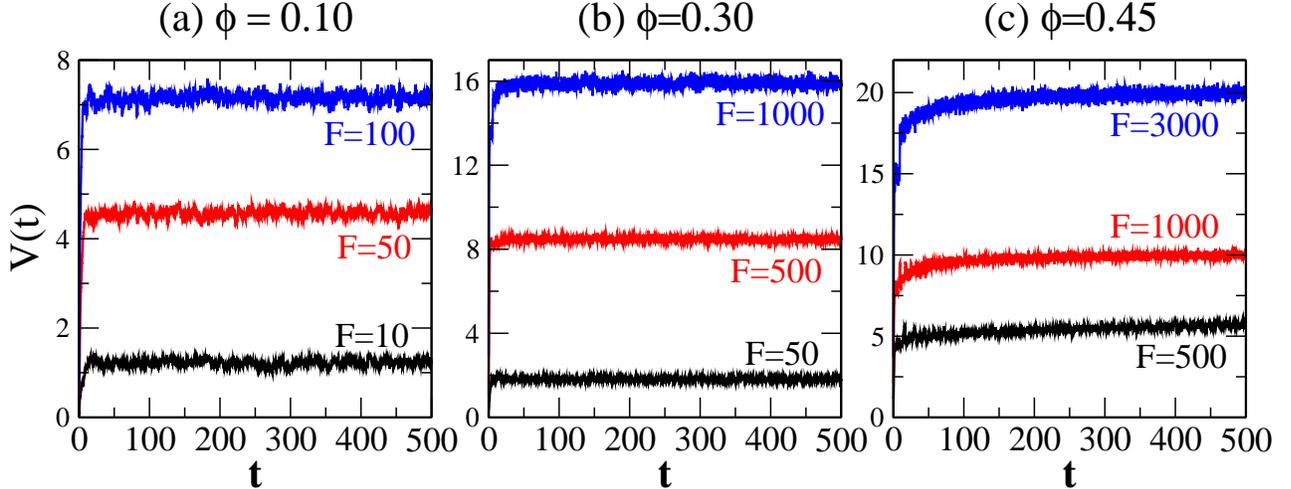}
	\caption{\label{fig2} The instantaneous velocity $V(t)$ of the intruder for different volume fractions: (a) $\phi=0.10$, (b) $\phi=0.30$, and (c) $\phi=0.45$. We plot $V(t)$ vs. $t$ for different values of $F$, as indicated. The diameter of the intruder hard-core is $R_{\rm int}=3R_{\rm 1}$.}
\end{figure}

In Fig.~\ref{fig2}, we plot the drift velocity of the intruder [$V(t)$ vs. $t$] for different values of $F$ and $\phi$. This is obtained as an average over 25 different runs. The intruder velocity is measured in the reference frame of the center of mass of the grains. For $\phi=0.10$ and very small $F(<0.1)$, the velocity of the intruder fluctuates around 0 over the simulation window. As we increase $F$, $V(t)$ grows and the intruder drifts along the direction of the force, as shown in Fig.~\ref{fig2}(a). At later times, the intruder acquires a steady velocity $V_s$. For $\phi=0.30$ [Fig.~\ref{fig2}(b)] and $\phi=0.45$ [Fig.~\ref{fig2}(c)], we observe a similar dependence of the drift velocity on time. Clearly, to achieve a given $V_s$, we need larger $F$ for denser systems.

Let us present some scaling arguments to understand the limiting behavior of the $V_s$-$F$ relationship. We first consider the case with small $\phi$ and $F$. In the steady state, the intruder is not accelerating, i.e., $F$ is balanced by the force applied by the grains on the intruder in the opposite direction. The latter force is proportional to the number of grains met by the intruder per unit time multiplied by the momentum transfer per event. It does not matter whether that event is an instantaneous collision, or a deformation event of longer duration. When $F$ is small, the intruder is slow and the momentum transfer is determined by the much faster, vibrated grains. As the number of particles met by the intruder per unit time is proportional to its velocity, we obtain
\begin{eqnarray}
\label{exp1}
F \sim \pi (R_{\rm int} + R_1)^2V_{\rm s} \phi \cdot m v_{\rm rms},
\end{eqnarray}
where $v_{\rm rms}$ is the root-mean-squared velocity of the grains. Eq.~(\ref{exp1}) yields $V_s \sim F$, which is identified as the Stokes law for particle motion through a viscous medium.

However, in the limit of large $F$, the intruder moves fast compared to the grains. Then, the momentum transfer on collision is proportional to $V_s$, which becomes the only relevant velocity scale in the problem. In this case,
\begin{eqnarray}
\label{exp2}
F \sim \pi (R_{\rm int} + R_1)^2V_{\rm s} \phi \cdot mV_{\rm s},
\end{eqnarray}
yielding the non-Stokes behavior, $V_s \sim F^{1/2}$. This result depends only on the fact that the velocity of the intruder is considerably larger than that of the grains. The Stokes $\rightarrow$ non-Stokes crossover occurs at $V_s^{\rm cross} \sim v_{\rm rms}$ (which is independent of $\phi$), and $F^{\rm cross} \sim \phi~v_{\rm rms}^2$.

\begin{figure}
	\centering
	\includegraphics*[width=0.93\textwidth]{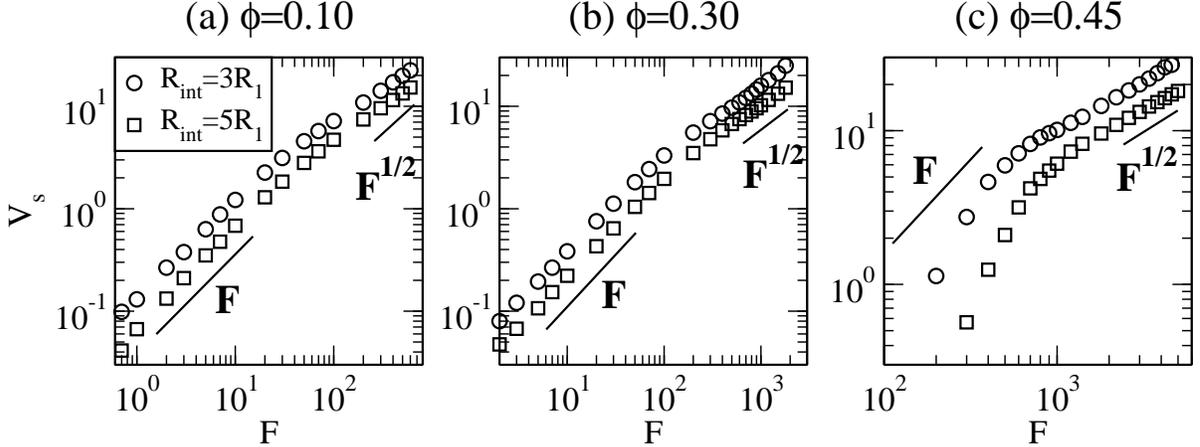}
	\caption{\label{fig3} Log-log plot of the steady-state velocity $V_s$ of the intruder vs. the driving force $F$ for different volume fractions: (a) $\phi=0.10$, (b) $\phi=0.30$, and (c) $\phi=0.45$. We show data for $R_{\rm int} = 3 R_1, 5 R_1$, denoted by the indicated symbols. The solid lines denote the limiting dependences: $V_s \sim F$ in the Stokes regime and $V_s \sim F^{1/2}$ in the non-Stokes regime, respectively.}
\end{figure}
Figure~\ref{fig3} shows the variation of $V_s$ with the external force $F$ for different $\phi$ and intruder sizes $R_{\rm int}$. (We obtain $V_s$ by time-averaging $V(t)$ in the steady state.) For $\phi = 0.10, 0.30$ [Figs.~\ref{fig3}(a)-(b)], we observe the Stokes regime $V_s \sim F$ for small forces, which crosses over to $V_s \sim F^{1/2}$ for large forces. It is clear from Figs.~\ref{fig3}(a)-(b) that $V_s^{\rm cross}$ is independent of $\phi$, and $F^{\rm cross}$ increases linearly with $\phi$, as argued above. In Fig.~\ref{fig3}(c), we plot $V_s$ vs. $F$ for $\phi=0.45$. We do not see a clear indication of the Stokes regime in this case. At these higher values of $\phi$, grains create a weak solid structure due to mutual overlap. Therefore, a critical force must be applied to the intruder to break this structure. Below the critical force, the intruder shows creep motion, giving rise to the non-Stokes behavior. For $F > F^{\rm cross}$, we again obtain $V_s \sim F^{1/2}$.

\begin{figure}
	\centering
	\includegraphics*[width=0.85\textwidth]{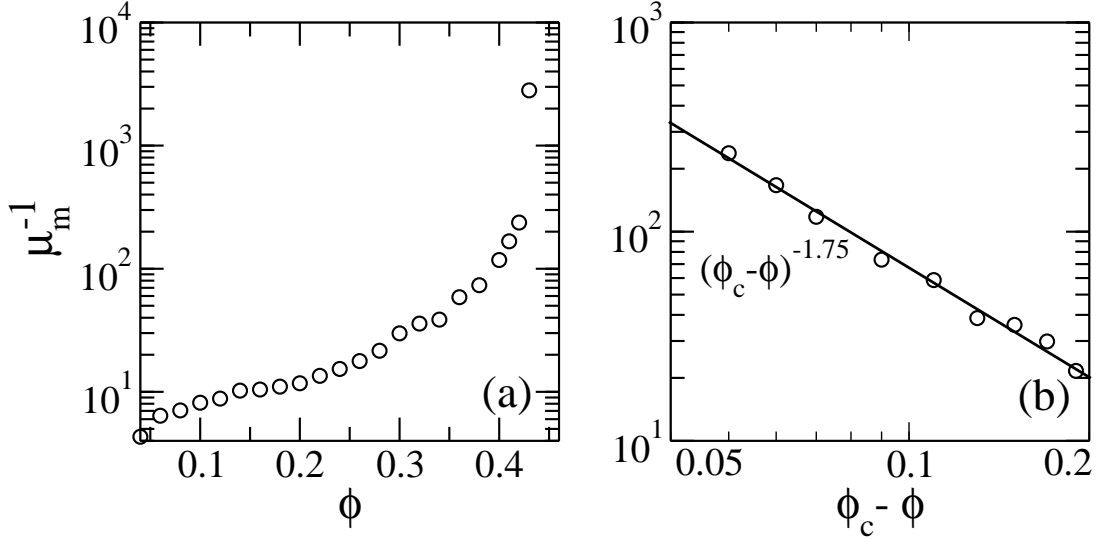}
	\caption{\label{fig4} Plot of inverse mobility of the intruder $\mu_m^{-1}$ as a function of the volume fraction $\phi$. We show data for intruder size $R_{\rm int} = 5 R_1$. (a) Linear-log plot of $\mu_m^{-1}$ vs. $\phi$, and (b) Log-log plot of $\mu_m^{-1}$ vs. $(\phi_J^{\rm eff} -\phi)$, with $\phi_J^{\rm eff}=0.48$. The solid line in (b) corresponds to the best power-law fit to our numerical data: $\mu_m^{-1} \sim (\phi_J^{\rm eff} -\phi)^{-1.75}$.}
\end{figure}

Next, we calculate the mobility of the intruder $\mu_m$ for different volume fractions of the grains. This quantity measures how easily the intruder can move with a constant velocity. It is defined as follows:
\begin{eqnarray}
\label{mob}
\mu_m= \frac{dV_s}{dF} \bigg|_{F=0},
\end{eqnarray}
and is proportional to the inverse viscosity of the granular system. We numerically obtain $\mu_m$ as the slope of the $V_s$-$F$ curve for $F \ll F^{\rm cross}$. In Fig.~\ref{fig4}, we plot $\mu_m^{-1}$ vs. $\phi$ for $R_{\rm int} = 5R_1$ and $\phi \in [0.04,0.45]$, i.e., more than a decade in density. Clearly, as we approach the jamming fraction of the grains, $\mu_m^{-1}$ diverges. Recall that $\phi_J \simeq 0.639$ for frictionless hard spheres. In our simulation, the grains are frictional and their outer radius is $R_2 = 1.1$. We estimate the effective jamming fraction as $\phi_J^{\rm eff} = \phi_J/R_2^3 \simeq 0.48$. We find that $\mu_m^{-1}$ increases with intruder size, as the number of grains interacting with the intruder becomes larger. Further, $\phi_J^{\rm eff}$ (the point of divergence of $\mu_m^{-1}$) is independent of the intruder size. Our data for $R_{\rm int} = 5R_1$ is consistent with $\mu_m^{-1} \sim (\phi_J^{\rm eff} -\phi)^{-\gamma}$, as shown in Fig.~\ref{fig4}(b). We obtain $\gamma \simeq 1.75$ from the best fit to our simulation data.

\begin{figure}
	\centering
	\includegraphics*[width=0.99\textwidth]{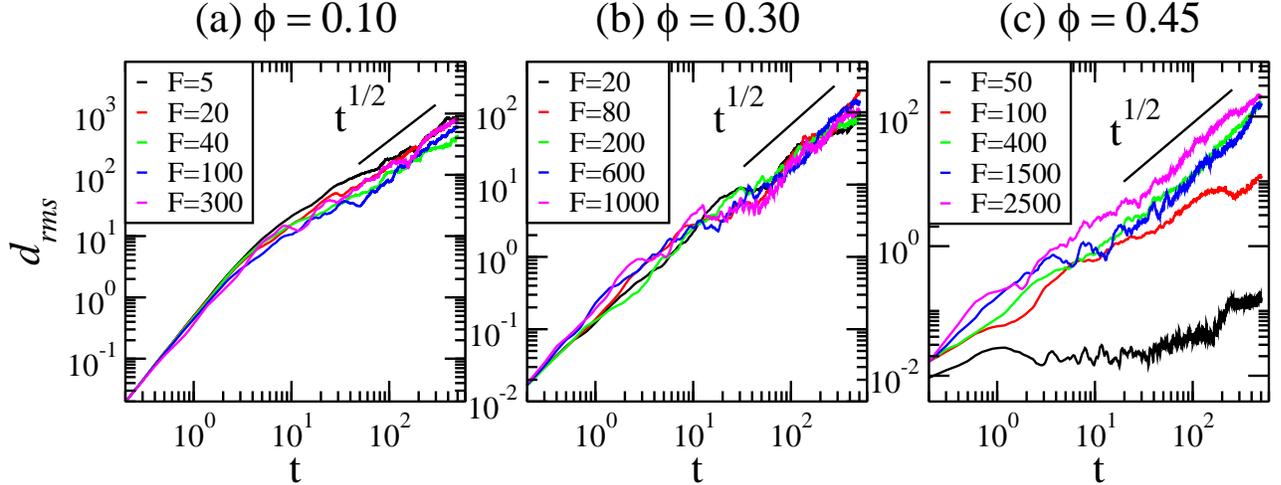}
	\caption{\label{fig5} Data for the root-mean-square displacement $d_{\rm rms}$ of the intruder (with hard-core diameter $R_{\rm int} = 3R_1$) in the $(y,z)$-plane. We plot $d_{\rm rms}$ vs. $t$ for different volume fractions: (a) $\phi=0.10$, (b) $\phi=0.30$, and (c) $\phi=0.45$.}
\end{figure}

To characterize the transverse motion of the intruder, we obtain the root-mean-squared displacement $d_{\rm rms} = \langle \vec{R}^2(t) \rangle^{1/2}$ in the $(y,z)$-plane. The angular brackets denote an averaging over independent runs. We plot $d_{\rm rms}$ vs. $t$ for $R_{\rm int} = 3R_1$ and different $\phi$ in Fig.~\ref{fig5}. For $\phi=0.10$ [Fig.~\ref{fig5}(a)], the intruder shows diffusive motion with $d_{\rm rms} \sim t^{1/2}$ after an initial transient regime for all values of $F$. We observed similar behavior for $\phi=0.30$, as shown in Fig.~\ref{fig5}(b). As expected, the initial transient regime becomes shorter as $\phi$ is increased.
For $\phi=0.45$ [Fig.~\ref{fig5}(c)], we see a strong dependence of the lateral distance traversed over a given time on the driving force. For small values of $F$, e.g., $F=50$, $d_{\rm rms}$ does not increase systematically. This small-force behavior is quite interesting and may have a number of explanations. We need better statistics to clarify this point. The data for $F=100$ shows some lateral motion, but it is intermediate between trapping and diffusion. For $F=400, 1500, 2500$, we find $d_{\rm rms}\sim t^{1/2}$ after a transient regime (similar to the low-density cases), as shown in Fig.~\ref{fig5}(c). Our data for $d_{\rm rms}$ vs. $t$ shows large fluctuations, especially at higher packing fractions. The improvement of this data would require substantial computational effort. Nevertheless, it is clear that the intruder exhibits Brownian motion in the lateral plane, as long as it is mobile. This result is independent of $\phi$ and $R_{\rm int}$.

In the present paper, we have focused on the motion of the intruder. It is equally interesting to study the behavior of the granular medium in the vicinity of the intruder~\cite{jewel,kozlow,carlev}, cf. Fig.~\ref{fig1}. There is a kinetic interplay between the motion of the intruder and the grains. In the context of Fig.~\ref{fig1}, we make the following observation about the cloud of grains surrounding the intruder. For relatively low applied forces, the velocities of the vibrated grains around the intruder appear random. As $F$ is increased, the grain velocities become more aligned with the direction of the force. The applicable picture is that a cloud of grains accompanies the intruder as it moves along the $+x$-direction. The properties of this cloud depend on $\phi$ and $F$. We do not pursue this point further in the present paper. However, it is intriguing enough to motivate a proper quantitative study of the dynamics of the cloud around the intruder. Further, it will be relevant to extend the present study to a group of intruders of different shapes, sizes, pulling speeds, etc. This would provide a better understanding of mutual interactions among intruders when they move through a granular medium~\cite{FPV2010,CD2013,MH2018,MSJ2018,AKR2020}.

\section{Summary and Discussion}
\label{sec4}

Let us conclude this paper with a summary and discussion of our results. We have studied the motion of an intruder through a granular medium by using large-scale molecular dynamics (MD) simulations in $d=3$. Our MD study incorporates both translational and rotational degrees of freedom. To the best of our knowledge, this is the first numerical study of this challenging problem. The energy-dissipation mechanism is solid friction between any pair of interacting particles (either intruder-grain or grain-grain). A constant external force $F$ is applied to the intruder. We also perturb the linear/angular velocities of the grains by using a white-noise thermostat, which keeps the density field of the granular medium homogeneous. In the absence of the thermostat, the granular material shows spontaneous dissipation-induced clustering~\cite{dp03,dap03,ap06,ap07,dps16,dps17}.

Our major results can be summarized as follows: \\
(a) The intruder velocity $V_s$ shows a power-law dependence on $F$ as $V_s \sim F^\beta$. For small $F$, $\beta = 1$, corresponding to the Stokes regime. For larger $F$, $\beta = 1/2$, corresponding to non-Stokes behavior. We have provided simple scaling arguments to understand both limits and the nature of the crossover. Our numerical results enable a systematic interpretation of a large variety of experimental results, which have reported diverse values of $\beta$. \\
(b) The inverse mobility $\mu_m^{-1}$, which is proportional to the viscosity, diverges as the volume fraction $\phi \rightarrow \phi_J^{\rm eff}$, where $\phi_J^{\rm eff}$ is the effective jamming fraction for our system. This divergence is consistent with a power-law behavior: $\mu_m^{-1} \sim (\phi_J^{\rm eff} - \phi)^{-\gamma}$, where $\gamma \simeq 1.75$. This power-law behavior is independent of the intruder size. \\
(c) After an initial transient regime, the intruder performs Brownian motion in the plane transverse to the direction of the external force.

In this paper, we have focused on the intruder dynamics, which has many interesting features. Clearly, the motion of the cloud of grains surrounding the intruder is also of great interest. We will tackle this problem in future work. We hope that the present study will motivate further experiments and simulations of the intruder problem. There is a pressing requirement for clean and unambiguous results for this problem, which could provide the basis for a better analytical understanding.

\subsubsection*{Acknowledgments}

PD acknowledges financial support from the Council of Scientific and Industrial Research, India. The research of MS, grant number 839/14, was supported by the ISF within the ISF-UGC joint research program framework.

\end{document}